\documentstyle[12pt]{article}
\def\appendix#1{
  \addtocounter{section}{1}
  \setcounter{equation}{0}
  \renewcommand{\thesection}{\Alph{section}}
 \section*{Appendix \thesection\protect\indent \parbox[t]{11.715cm} {#1}}
  \addcontentsline{toc}{section}{Appendix \thesection\ \ \ #1}
  }

\newcommand{\newsection}{
\setcounter{equation}{0}
\section}

\newcommand{\tr}[1]{\:{\rm tr}\,#1}
\def\const{{\rm const}}
\def\e{{\,\rm e}\,}
\def\Evac{E_{{\rm vac}}}

\def\H{H_{{\rm YM}}}
\def\vac{|0\rangle}
\newcommand{\rf}[1]{(\ref{#1})}
\newcommand{\non}{\nonumber \\*}
\renewcommand{\l}[1]{\left.\frac{\partial }{\partial
\lambda }#1\right|_{\lambda =0}}
\begin{document}
\begin{titlepage}
\begin{flushright}
ITEP--TH--34/98\\
hep-th/9806150\\
June, 1998
\end{flushright}
\vspace{1.5cm}

\begin{center}
{\LARGE Interquark Potential in Schr\"odinger Representation}
\\[.5cm]
\vspace{1.9cm}
{\large K.~Zarembo}\\
\vspace{24pt}
{\it Department of Physics and Astronomy,}
\\{\it University of British Columbia,}
\\ {\it 6224 Agricultural Road, Vancouver, B.C. Canada V6T 1Z1}
\\ \vskip .2 cm
and\\ \vskip .2cm
{\it Institute of Theoretical and Experimental Physics,}
\\ {\it B. Cheremushkinskaya 25, 117259 Moscow, Russia} \\ \vskip .5 cm
E-mail: {\tt zarembo@itep.ru/@theory.physics.ubc.ca}
\end{center}
\vskip 2 cm
\begin{abstract}
Static charges are introduced in Yang-Mills theory via coupling to
heavy fermions. The states containing static color charges are constructed
using integration over gauge transformations. A functional
representation for interquark potential is obtained. This representation
provides a simple criterion for confinement.
\end{abstract}

\end{titlepage}

\setcounter{page}{2}

\newsection{Introduction}

The gluon vacuum is expected to determine many
properties of the low-energy QCD. According to the most straightforward
definition, the vacuum is described by a ground state
wave function -- the lowest energy solution of the stationary
Schr\"odinger equation. The exact ground state in gluodynamics must be
very complicated, but one may hope that a reasonably good approximation
can be obtained by means of variational or of some other approximate
method if a proper parametrization of the wave functional is used. Such
parametrization should respect the symmetries of the Yang-Mills theory,
first of all, the gauge invariance. There is a simple way to achieve
the invariance of a wave functional -- to enforce it by
integrating over all gauge transformations
\cite{kk94,zar98,hmvi98,dia98}:
\begin{equation}\label{ansatz}
\Psi [A]=\int [DU]\,\e^{-S[A^U]},
\end{equation}
\begin{equation}\label{AU}
A_i^U =
U ^{\dagger}\left(A_i+\frac{1}{g}\,\partial
_i\right)U .
\end{equation}
Alternative approaches are based on the
use of gauge-invariant variables from the very beginning \cite{giv} or on
fixing of residual gauge symmetry \cite{gf}.

The representation of a projector on the physical states in
a form of the integral over gauge transformations is well known in
Yang-Mills theory at finite temperature \cite{ft}. In the present
context this method appears to be very convenient for variational
calculations, since it allows to make the
simplest, Gaussian variational ansatz \cite{kk94,hmvi98,dia98}
exactly gauge invariant.

The averaging over
the gauge group introduces additional variables, the parameters of gauge
transformations, which more or less correspond to longitudinal
components of the gluon fields.  Introduction of such variables might seem
somewhat artificial, since longitudinal degrees of freedom do not
propogate and, thus, are unphysical in pure gauge theory. This is true,
however, until external color charges are considered. Electric fields of
a static charge are longitudinal, so longitudinal components of gauge
fields are responsible for color electric forces, which, in their turn,
are responsible for confinement.
Because averaging over gauge
transformations explicitly separates longitudinal degrees of freedom, it
provides an appropriate framework for consideration of static color
charges and of their interaction \cite{zar98}. In this paper the
averaging over gauges is applied to Yang-Mills theory coupled to heavy
fermions, which allows to introduce static charges in a more systematic
way.

\newsection{Coupling to heavy fermions}

The ground state in gluodynamics satisfies Schr\"odinger equation
\begin{equation}\label{schr}
\H\Psi =\Evac\Psi ,
\end{equation}
where $\H$ is Yang-Mills Hamiltonian (in $A_0=0$ gauge):
\begin{equation}\label{ham}
\H=\int d^3x\,\left(\frac{1}{2}\,E_i^AE_i^A
+\frac{1}{4}\,F_{ij}^AF_{ij}^A\right).
\end{equation}
When the wave function is a functional of
the gauge potentials, the electric fields act as variational derivatives:
\begin{equation}\label{elect}
E_i^A=-i\,\frac{\delta }{\delta A_i^A}.
\end{equation}
We consider $SU(N)$ gauge group with generators
$T^A$ normalized by $\tr T^AT^B=-\delta ^{AB}/2$ and sometimes use matrix
notations for gauge potentials:  $A_i=A_i^AT^A$.

The Hamiltonian \rf{ham} commutes with operators $D_iE_i^A$, and the
ground state wave functional is subject to the Gauss' law constraint:
\begin{equation}\label{gauss}
D_iE_i^A\, \Psi =0.
\end{equation}
This constraint represents in the infinitesimal form the gauge
invariance of the ground state\footnote{For simplicity,
we disregard peculiarities
related to topologically nontrivial gauge transformations, see
\protect\cite{jac80}.}:
\begin{equation}\label{ginv} \Psi [A^\Omega]=\Psi [A].
\end{equation}
The integral \rf{ansatz} can be considered as
a general solution of the Gauss' law constraint. Any gauge-invariant
state can be represented in this form, the exact ground state in
gluodynamics, in particular. Of course, the representation
\rf{ansatz} is aimed to the use of approximate methods, when some
simple form of the action $S[A]$ is employed. In the variational
approach of Ref.~\cite{kk94} the action was taken quaratic and
diagonal in spatial indices. This variational ansatz was generalized 
to include longitudinal
part of the quadratic form \cite{dia98} and tree-level condensates
\cite{hmvi98}. However, since the consideration below is purely
kinamatical, we do not restrict ourselves 
to any particular form of the bare wave functional $\exp(-S[A])$.

In order to introduce color charges, we couple the gauge fields to
heavy fermions, so heavy that it is possible to neglect their kinetic
term in the Hamiltonian:
\begin{equation}\label{hamf}
H=\int d^3x\,\left(\frac{1}{2}\,E_i^AE_i^A
+\frac{1}{4}\,F_{ij}^AF_{ij}^A+M\bar{\psi }\psi \right).
\end{equation}
The fermion operators obey anticommutation relations
\begin{equation}\label{ac}
\left\{\psi^{a}_\alpha(x),
\psi ^{\dagger}_{\beta b}(y)\right\}
=\delta _{\alpha \beta }\delta^{a}_{\phantom{a}b}\delta (x-y),
\end{equation}
where $a,b$ and $\alpha ,\beta $ are color and spinor indices,
respectively. The fermion fields transform in the fundamental
representation of $SU(N)$:
\begin{equation}\label{psiu} \psi
^U=U^{\dagger}\psi,~~~~~ \psi ^{\dagger\,U}=\psi^{\dagger}U.
\end{equation}
Physical states are gauge invariant and satisfy the Gauss' law
constraint:
\begin{equation}\label{gausshf}
\left(D_iE_i^A+i\,\psi ^{\dagger}T^A\psi \right)\, \Psi_{{\rm phys}} =0.
\end{equation}

Because of the absence of the kinetic term for matter fields,
fermion operators
obey simple commutation relations with the Hamiltonian. In the standart
representation of the Dirac matrices, when $\gamma ^0$ is diagonal:
\begin{equation}\label{gamma0}
\gamma ^0=\left(
\begin{array}{cc}
{\bf 1}&0\\
0&{\bf -1}\\
\end{array}
\right),
\end{equation}
the commutation relations take the form:
\begin{eqnarray}\label{hpsi}
\left[H,\psi _\alpha (x)\right] &=& \pm M\psi _\alpha (x),
~~~~~\left\{
\begin{array}{lr}
+,&\alpha =3,4\\
-,&\alpha =1,2\\
\end{array}\right.
\\*
\left[H,\psi^{\dagger}_\alpha (x)\right] &=&
\pm M\psi^{\dagger}_\alpha  (x),
~~~~~\left\{
\begin{array}{lr}
+,&\alpha =1,2\\
-,&\alpha =3,4\\
\end{array}\right.\,.
\end{eqnarray}
Therefore, $\psi _{3,4}(x), \psi ^{\dagger}_{1,2}(x)$ and
$\psi _{1,2}(x), \psi ^{\dagger}_{3,4}(x)$ are creation and annihilation
operators, respectively. The fermion vacuum is defined by equations
\begin{eqnarray}\label{vacuum}
\psi _{1,2}(x)\vac&=&0,
\non
\psi^{\dagger} _{3,4}(x)\vac&=&0.
\end{eqnarray}
The fermion vacuum is gauge invariant by itself, and the ground state of
the Hamiltonian \rf{hamf} is described by the wave function
\begin{equation}\label{gst}
\Psi _{{\rm vac}}=\Psi [A]\vac=\int [DU]\,\e^{-S[A^U]}\,\vac,
\end{equation}
where $\Psi [A]$ is the vacuum wave functional in gluodynamics -- the
lowest-energy solution of the Schr\"odinger equation \rf{schr}.

Excited fermion states are obtained by acting of creation operators on the
vacuum. These states are no more gauge invariant, and it is necessary to
average them over gauge transfromation in order to project on the
physical subspace. The physical states are, consequently, constructed
with the help of dressed operators
\begin{eqnarray}\label{dressed}
\psi ^{U\,a'}_\alpha(y)&=&
U^{\dagger\,a'}_{\phantom{\dagger\,a'}a}(y)
\psi^a_{\alpha }(y),
\non
\psi ^{\dagger\,U}_{\beta b'}(x)
&=&\psi ^{\dagger}_{\beta b}(x)
U^b_{\phantom{b}b'}(x),
\end{eqnarray}
where $\alpha =3,4$ and $\beta =1,2$ (below spinor indices are omitted
for brevity). The most general state of this type,
\begin{eqnarray}\label{general}
\Psi _{a'_1\ldots a'_n}^{\phantom{a'_1\ldots a'_n}b'_1\ldots b'_s}
(x_1,\ldots,x_n,y_1,\ldots,y_s)
&=&\int [DU]\,\e^{-S[A^U]}\,
\psi^{\dagger\,U}_{a'_1}(x_1)\ldots
\psi^{\dagger\,U}_{a'_n}(x_n)
\non &&
\psi ^{U\,b'_1}(y_1)\ldots\psi^{U\,b'_s}(y_s)\vac,
\end{eqnarray}
describes $n$ static antiquarks and $s$ quarks located at points
$x_i$, $y_i$. Dressed operators \rf{dressed} are invariant under local
gauge transformations because gauge rotations of the gluon and of the
quark fields; $A_i\rightarrow A_i^\Omega $, 
$\psi \rightarrow\psi ^\Omega$,
$\psi ^{\dagger}\rightarrow\psi ^{\dagger\,\Omega }$; are compensated
by a change of variables $U^a_{\phantom{a}a'}\rightarrow\Omega
^{\dagger\,a}_{\phantom{\dagger\,a} b}U^b_{\phantom{b}a'}$ in the
integral over the gauge group. These transfromations do not act on
indices with prime. To be more presize, this statement is true for
transformations with $\lim\limits_{x\rightarrow\infty}\Omega(x)=1$. If
$\lim\limits_{x\rightarrow\infty}\Omega(x)=\Omega _\infty \neq 1$,
the change of variables $U\rightarrow \Omega ^{\dagger}U$ violates
boundary conditions in the path integral over $U$ and should be
accomponied by a global transformation $U\rightarrow U\Omega_\infty$.
This means that indices with prime are global color ones.

The state \rf{general} can be factorized in a product of the gluon and
the fermion wave functions. For example,
quark-antiquark pair in a color-singlet state is described by the wave
function
\begin{equation}\label{mes}
\Psi _M(x,y)=\int [DU]\,\e^{-S[A^U]}\,
\psi^{\dagger\,U}_{a'}(x)
\psi ^{U\,a'}(y)\vac
=\Psi^a_{\phantom{a}b}[A;x,y]\,
\psi^{\dagger}_{a}(x)\psi ^{b}(y)\vac,
\end{equation}
where
\begin{equation}\label{qq}
\Psi^a_{\phantom{a}b}[A;x,y]
=\int [DU]\,\e^{-S[A^U]}\,
U^a_{\phantom{a}a'}(x)U^{\dagger\,a'}_{\phantom{\dagger\,a'}b}(y).
\end{equation}
The gluon wave function \rf{qq} transforms in the representation
$\bar{N}\otimes N$ of the gauge group:
\begin{equation}\label{NxN}
\Psi^a_{\phantom{a}b}[A^\Omega ;x,y]
=\Omega ^{\dagger\,a}_{\phantom{\dagger\,a}c}(x)
\Omega^d_{\phantom{d}b}(y)
\Psi^c_{\phantom{c}d}[A;x,y]
\end{equation}
and satisfies Schr\"odinger equation for pure Yang-Mills theory.

Again, any state obeying transformation law \rf{NxN} can be
represented in the form of an integral \rf{qq}; moreover,
the same action $S[A]$ can be
used both in the vacuum and in the charged sectors. The integration over
gauge transformations projects the bare wave functional $\exp(-S[A])$ on
subspaces of states with definite transformation properties, gauge
invariant states in the former case and
the ones belonging to $\bar{N}\otimes N$ representation in the latter.
These subspaces are orthogonal in the full Hilbert space, and it is
always possible to find a state which has given projections on each
of them. Hence, the action $S[A]$ can be chosen in such a way
that the projections of the wave functional $\exp(-S[A])$ on the
sectors defined by transfromation laws \rf{ginv} and \rf{NxN} are the
lowest-energy eigenstates of the Yang-Mills Hamiltonian in each 
sector.  So, the vacuum, the quark-antiquark and, in fact, any state of
type \rf{general} can be obtained by averaging
of the same bare wave functional over gauge transformations.
The above arguments concern exact wave
functionals, but the statement that one can use the same action $S[A]$
in different sectors is also true within a
variational approach. Suppose that the form of $S[A]$ is fixed 
up to some parameters, then
variational estimates for these parameters will be the same
in the vacuum and in the charged sectors \cite{zar98} because energies of
the ground states in these sectors, being proportional to the
volume, differ by an amount which is finite in the infinite volume limit
and, thus, does alter variational equations.

Another example of the state \rf{general} is the one with baryonic
quantum numbers:
\begin{eqnarray}\label{bar}
\Psi _B(y_1,\ldots,y_N)&=&\int [DU]\,\e^{-S[A^U]}\,
\varepsilon _{a'_1\ldots a'_N}\psi ^{U\,a'_1}(y_1)\ldots
\psi ^{U\,a'_N}(y_N)\vac
\non
&=&\Psi_{a_1\ldots a_N}[A;y_1,\ldots,y_N]\,
\psi ^{a_1}(y_1)\ldots\psi ^{a_N}(y_N)\vac,
\end{eqnarray}
\begin{equation}\label{barglue}
\Psi_{a_1\ldots a_N}[A;y_1,\ldots,y_N]
=\int [DU]\,\e^{-S[A^U]}\,\varepsilon _{a'_1\ldots a'_N}
U^{\dagger\,a'_1}_{\phantom{\dagger\,a'_1}a_1}(y_1)\ldots
U^{\dagger\,a'_N}_{\phantom{\dagger\,a'_N}a_N}(y_N).
\end{equation}

The state \rf{qq} has simple physical meaning, it describes color
fields of static quark-antiquark pair in the gluon vacuum.
In the Abelian theory, Coulomb field of a static charge is created by an
operator $\e^{ieV(x)}$  \cite{dir55}, where
\begin{equation}\label{defV}
V(x)=\int d^3y\,\frac{1}{-\partial ^2}(x-y)\partial _iA_i(y),
\end{equation}
so dressed electron operators have the form \cite{dir55}:
\begin{eqnarray}\label{dressedab}
\psi ^{(*)}_\alpha(y)&=&\e^{-ieV(y)}
\psi_{\alpha }(y),
\non
\psi ^{(*)\,\dagger}_{\beta}(x)
&=&\e^{ieV(x)}\psi ^{\dagger}_{\beta}(x).
\end{eqnarray}
It is not difficult to reproduce this construction from the
representation based on averaging over gauge transformations. In Abelian
theory $U=\e^{ie\varphi }$, $A^U_i=A_i+\partial _i\varphi$ and the action
$S[A]$ should be
 quadratic; for simplicity we choose the diagonal quadratic form.
The integration over gauge transformations in \rf{qq} is Gaussian and can
be done explicitly:
\begin{eqnarray}\label{abel}
\Psi [A;x,y]&=&\int [d\varphi ]\,\exp\left[-\frac{1}{2}\int
d^3xd^3w\,\left(A_i+\partial _i\varphi \right)(z)\,
K(z-w)\,\left(A_i+\partial _i\varphi \right)(w)
\right.
\non &&\left.\vphantom{\frac{1}{2}}+
ie\varphi (x)-ie\varphi (y)\right]
\non
&=&C\,\e^{ieV(x)}\e^{-ieV(y)}
\exp\left(-\frac{1}{2}\int d^3zd^3w\,A_i^\bot(z)K(z-w)A_i^\bot(w)
\right),
\end{eqnarray}
where $C$ is a field-independent constant and $A_i^\bot(z)$ denotes
transversal part of the gauge potentials:
\begin{equation}\label{trans}
A_i^\bot=\left(\delta
_{ij}-\frac{\partial _i\partial _j}{\partial ^2}\right)A_j.
\end{equation}
As follows from conformal symmetry, the coefficient function $K$ is equal
to $|p|$ in the momentum space. Then the last factor in \rf{abel} is
nothing but the
vacuum wave functional for free electro-magnetic field. The first two
factors reproduce dressing operators which correspond to a pair of
static charges of opposite sign located at $x$ and $y$.
 More generally, the integration over the Abelian gauge group
in \rf{general}, due to its Gaussian nature, leads to the replacement of
$\psi ^{U}$ by $\psi ^{(*)}$, so dressing \rf{dressed},
\rf{general} is equivalent to
\rf{dressedab} in the Abelian theory.

Dressed operators \rf{dressedab} possess a number of important
properties. They are gauge-invariant and create eigenstates of the free
electro-magnetic Hamiltonian. In QED they also play an important role,
because dressed electron and positron operators allow to avoid IR
divergencies related to emission
of soft photons \cite{lm95,ir}. Different non-Abelian generalazations of
the operators \rf{dressedab} were proposed \cite{lm95,dr}.  In contrast
with these constructions, dressing \rf{general} in non-Abelian case does
not have factorized form, the integral over gauge transformations is no
more Gaussian, and the operator creating static gauge fields of one
particular charge cannot be defined; that is, the form of the wave
functional describing this charge depends on the presence of other
charges.

\newsection{Interquark potential}

The energy of the state \rf{mes} after obvious subtractions determines
$q\bar{q}$ interaction potential:
\begin{equation}\label{e12}
\frac{\left\langle\Psi_{M}|H|\Psi_{M}\right\rangle}
{\left\langle\Psi_{M} |\Psi_{M} \right\rangle}
=2M+\Evac+V(x-y).
\end{equation}
Matrix elements of gauge-invariant states, like those entering
eq.~\rf{e12}, are proportional to the volume of the gauge group. The
representation \rf{ansatz} allows to get rid of this infinite factor
easily \cite{kk94}.  For example, the norm of the vacuum state is
\begin{equation}\label{norm}
\left\langle\Psi |\Psi \right\rangle=\int
[DU][DU'][dA]\, \e^{-S[A^U]-S[A^{U'}]}=
\const\,\int [DU][dA]\,\e^{-S[A^U]-S[A]}.
\end{equation}
The norm of the charged state \rf{mes} is
\begin{equation}\label{norm'}
\left\langle\Psi_{M} |\Psi_{M} \right\rangle
=\const\,\int [DU][dA]\,\tr U(x)U^{\dagger}(y)\e^{-S[A^U]-S[A]}.
\end{equation}
Matrix elements of the Hamiltonian correspond to insertion of the
operator $R[A]$,
\begin{equation}\label{defr}
R[A]=\int d^3x\,\left(\frac{1}{2}\,\frac{\delta
^2S[A]}{\delta A_i^A\delta
A_i^A}-\frac{1}{2}\,\frac{\delta S[A]}{\delta A_i^A}
\,\frac{\delta S[A]}{\delta A_i^A}
+\frac{1}{4}\,F_{ij}^AF_{ij}^A\right),
\end{equation}
if variational derivatives in the Hamiltonian act on the right. If
they act on the left, $R[A]$ is replaced by $R[A^U]$. It is convenient to
introduce a source for the symmetric combination $(R[A^U]+R[A])/2$:
\begin{equation}\label{defz}
Z=\int
[DU][dA]\,\e^{-S[A^U]-S[A]+\frac{\lambda }{2}\,R[A^U]+\frac{\lambda
}{2}\,R[A]},
\end{equation}
then
matrix elements of the Hamiltonian are obtained by differentiation with
respect to $\lambda $. For example, the vacuum energy is given by
\begin{equation}\label{evac}
\Evac=\l{\ln Z}.
\end{equation}
This quantity is proportional to the volume and contains UV divergent
contribution from zero-point oscillations. Regularized energy density can
be related to gluon condensate \cite{SVZ79}:
\begin{equation}\label{edens}
\Evac=\frac{\beta (\alpha _s)}{16\alpha _s}\left\langle
0\left|F_{\mu \nu }^AF^{A\mu \nu}\right|0\right\rangle
\,{\rm Vol}+\mbox{UV divergent terms},
\end{equation}
where $\beta (\alpha _s)$ is a $\beta $-function in Yang-Mills theory.

For $q\bar{q}$ potential we obtain:
\begin{equation}\label{v12}
V(x-y)=\l{\ln\left\langle\tr U(x)U^{\dagger}(y)\right\rangle},
\end{equation}
where averaging is defined by the partition function \rf{defz}. It is 
worth explaining how confinement can arise in such representation.
The confining potential grows with distance; at the same time, the
correlator in \rf{v12} should decrease 
at infinity. But its logarithm can
increase thus leading to the confining potential. It happens when
the integration over the gauge group produces a mass gap:
\begin{equation}\label{expfall}
\left\langle\tr U(x)U^{\dagger}(y)\right\rangle
\propto\e^{-mr},
\end{equation}
where $r=|x-y|\rightarrow\infty$. Then interquark potential grows
linearly with distance:
\begin{equation}\label{Vqq}
V(r)=\sigma r+\ldots,
\end{equation}
the string tension being equal to the derivative of the mass:
\begin{equation}\label{strten}
\sigma =-\left.\frac{\partial m}{\partial \lambda}\right|_{\lambda =0}.
\end{equation}
Hence, the confinement arises due to generation of a mass gap in the
averaging over gauge transformations. More presizely, the confining
potential is generated if there is a nonzero linear response of the mass
gap to switching of the Hamiltonian.

Analogously, the energy of quarks in the baryonic state \rf{bar},
\begin{equation}\label{vbar}
V_B(y_1,\ldots,y_N)=
\frac{\left\langle\Psi_{B}|H|\Psi_{B}\right\rangle}
{\left\langle\Psi_{B} |\Psi_{B} \right\rangle}
-NM-\Evac,
\end{equation}
can be represented as the derivative of
certain correlation function in the
statistical system defined by the partition function \rf{defz}:
\begin{equation}\label{v1...n}
V_B(y_1,\ldots,y_N)=\l{\ln\left\langle\varepsilon _{a_1\ldots a_N}
\varepsilon ^{b_1\ldots b_N}
U^{\dagger\,a_1}_{\phantom{\dagger\,a_1}b_1}(y_1)\ldots
U^{\dagger\,a_N}_{\phantom{\dagger\,a_N}b_N}(y_N)\right\rangle}.
\end{equation}

\newsection{Discussion}

Examples of calculations of interquark potential starting from
representation based on averaging over gauge transformations
are given in \cite{zar98,dia98}. These calculations use particular
assumptions about the structure of the ground state
wave functional in gluodynamics and are by no means derivations from first
principles. On the level of kinematical consideration, one can state
that the potential is confining if the integration
over gauge transformations generates a mass gap with nonzero linear
response on switching of the Hamiltonian. The calculations
\cite{zar98,dia98} show that generation of a mass gap
 is rather natural in
non-Abelian theory, but quantitative results can be obtained
only if the vacuum wave functional is known in some approximation.
Perhaps, the simplest Gaussian ansatz averaged over gauge
transformations \cite{kk94,dia98,hmvi98} can provide a reasonable
variational approximation. The main problem is correct renormalization.
The variational method always overestimates the vacuum energy, if
non-renormalized divergencies remain, the overestimate is infinite, and
variational approximation loses sense \cite{tik}. It was argued
\cite{dia98} that Gaussian ansatz can be made compatible with
asymptotic freedom; the renormalization properties of Yang-Mills
theory in the Schr\"odinger representation were also discussed in
\cite{ren} from different points of view.

\subsection*{Acknowledgments}

The author is grateful to I.~Kogan, M.~Polikarpov and G.~Semenoff for
discussions. This work was supported by NATO Science Fellowship and, in
 part, by CRDF grant 96-RP1-253, INTAS grant 96-0524, RFFI grant
 97-02-17927 and grant 96-15-96455 of the support of scientific schools.


\begin{thebibliography}{99}
\addtolength{\itemsep}{-6pt}

\bibitem{kk94}
I.I.~Kogan and A.~Kovner, Phys. Rev. D52 (1995) 3719, hep-th/9408081.

\bibitem{zar98}
K. Zarembo, Phys. Lett. B421 (1998) 325, hep-th/9710235.

\bibitem{hmvi98}
C. Heinemann, C. Martin, D. Vautherin and E. Iancu, hep-th/9802036.

\bibitem{dia98}
D. Diakonov, hep-th/9805137.

\bibitem{giv}
J. Goldstone and R. Jackiw, Phys. Lett. B74 (1978) 81;\\
Yu.A. Simonov, Sov. J. Nucl. Phys. 41 (1985) 835 
[Yad. Fiz. 41 (1985) 1311];\\ 
M. Bauer, D.Z. Freedman and P.E. Haagensen, Nucl. Phys. B428 (1994) 147,
hep-th/9405028;\\
P. E. Haagensen and K. Johnson, Nucl. Phys. B439 (1995) 597, 
hep-th/9408164;\\
D. Karabali and V.P. Nair, Nucl. Phys. B464 (1996) 135, 
hep-th/9510157;\\ 
P. E. Haagensen, K. Johnson and C. S. Lam, Nucl. Phys. B477 (1996) 273,
 hep-th/9511226. 

\bibitem{gf}
N.H. Christ and T.D. Lee, Phys. Rev. D22 (1980) 939;\\
P. van Baal, hep-th/9511119; hep-th/9711070.

\bibitem{ft}
A.M.~Polyakov, Phys. Lett. B72 (1978) 477;
{\it Gauge Fields and Strings} (Harwood Academic
Publishers, 1987);\\
L.~Susskind, Phys. Rev. D20 (1979) 2610;\\
B.~Svetitsky, Phys. Rep. 132 (1986) 1.

\bibitem{jac80}
R. Jackiw, Rev. Mod. Phys. 52 (1980) 661.

\bibitem{dir55}
P.A.M. Dirac, Can. J. Phys. 33 (1955) 650; {\it The Principles of Quantum
Mechanics} (Oxford, Clarendon Press, 1958).

\bibitem{lm95}
M. Lavelle and D. McMullan, Phys. Rep. 279 (1997) 1, hep-ph/9509344.

\bibitem{ir}
E. Bagan, M. Lavelle and D. McMullan, Phys. Rev. D56 (1997) 3732, 
hep-th/9602083; D57 (1998) 4521, hep-th/9712080. 

\bibitem{dr}
P.E. Haagensen and K. Johnson, hep-th/9702204;\\
M. Lavelle and D. McMullan, hep-th/9805013.

\bibitem{SVZ79}
M.~Shifman, A.~Vainshtein and V.~Zakharov, Nucl. Phys. B147 (1979)
385.

\bibitem{tik}
G. Tiktopoulos, Phys. Rev. D57 (1998) 6429, hep-th/9705230.

\bibitem{ren}
W.E.~Brown and I.I.~Kogan, hep-th/9705136;\\
K. Zarembo, hep-th/9803237; hep-ph/9804276.

\end{thebibliography}
\end{document}